\documentclass{article}

\usepackage{arxiv}

\usepackage[utf8]{inputenc} 
\usepackage[T1]{fontenc}    
\usepackage{hyperref}       
\usepackage{url}            
\usepackage{booktabs}       
\usepackage{amsmath}
\usepackage{nicefrac}       
\usepackage{microtype}      
\usepackage{graphicx}
\usepackage{doi}

\title{Cardiac reentry modeled by spatiotemporal chaos in a coupled map lattice}

\author{ {R. V. Stenzinger}\\
	Departamento de Física\\
	Universidade Federal de Santa Catarina\\
	Florian{\'o}polis, Brazil \\
	\texttt{rafaelvste@gmail.com} \\
	\And
	{M. H. R. Tragtenberg} \\
	Departamento de Física\\
	Universidade Federal de Santa Catarina\\
	Florian{\'o}polis, Brazil \\
	\texttt{marcelotragtenberg@gmail.com} \\
}

\date{}

\renewcommand{\headeright}{}
\renewcommand{\undertitle}{}

\hypersetup{
pdftitle={Cardiac reentry modeled by spatiotemporal chaos in a coupled map lattice},
pdfsubject={q-bio.NC, math.DS, physics.bio-ph},
pdfauthor={R. V. Stenzinger, M. H. R. Tragtenberg},
pdfkeywords={First keyword, Second keyword, More},
}

\begin{document}
\maketitle

\begin{abstract}
Arrhythmias are potentially fatal disruptions to the normal heart rhythm, but their underlying dynamics is still poorly understood. Theoretical modeling is an important tool to fill this gap. Typical studies often employ detailed multidimensional conductance-based models. We describe the cardiac muscle with a three-dimensional map-based membrane potential model in lattices. Although maps retain the biophysical behavior of cells and generate computationally efficient tissue models, few studies have used them to understand cardiac dynamics. Our study captures healthy and pathological behaviors with fewer parameters and simpler equations than conductance models. We successfully generalize results obtained previously with reaction-diffusion systems, showing how chaotic properties result in reentry, a pathological propagation of stimuli that evolves to arrhythmias with complex spatiotemporal features. The bifurcation diagram of the single cell is very similar to that obtained in a detailed conductance-based model. We find \textit{torsade de pointes}, a clinical manifestation of some types of tachycardia in the electrocardiogram, using a generic sampling of the whole network during spiral waves. We also find a novel type of dynamical pattern, with wavefronts composed of synchronized cardiac plateaus and bursts. Our study provides the first in-depth look at the use of map-based models to simulate complex cardiac dynamics.
\end{abstract}


\section{Introduction}
\label{sec:intro}

Spatiotemporal patterns are known to happen in biological tissue~\cite{Cross1993,Haken2008}. An earlier model is Alan Turing reaction-diffusion system for morphogenesis~\cite{Turing1952}. On another biological context, patterns can form through waves of electrical activity and take part in important processes, such as cardiac arrhythmias~\cite{Fenton2008,Alonso2016}. Some types of tachycardia, when the heart beats at abnormally high rates, can be understood as a high frequency spiral wave in the myocardium, following the interpretation of arrhythmias as rotating waves put forth by Wiener and Rosenblueth in two dimensions~\cite{Wiener1946} and by Winfree in three dimensions~\cite{winfree2001,Fenton2008}. During fibrillation, the heart muscle is overtaken by multiples wavefronts of excitation, such as smaller spirals. Fibrillation can evolve from tachycardia and lead to sudden-death. As many patients who die suddenly do not have any identifiable signs of high-risk, understanding the evolution and transition of different cardiac rhythms is of great importance to prevent such casualties~\cite{Glass2005,Weiss2010}.

Pattern-forming arrhythmias begin with reentry, when the cardiac muscle is re-excited by a stimulus out of its normal route. An example is a patch of cardiac cells that form a conduction block by means of refractory period, but recover excitability right after the passage of a stimulus nearby, leading the wave to propagate "backwards" into the recovered region. This configuration often evolves to complex spatiotemporal patterns and spiral waves. At the cellular level, a mechanism to induce reentry is triggered activities~\cite{Fenton2008}, where oscillations in the membrane potential are triggered by the preceding action potential. These oscillations are known as afterdepolarizations and are of two types: early (EAD) and delayed (DAD). EAD happens during the plateau or recovering phase of the cardiac action potential and DAD during the resting state~\cite{Gaztanaga2012}. EAD is the result of the increase of a depolarizing ionic current, the reduction of a repolarizing one or both~\cite{Weiss2010}.

A mechanism that leads to EAD occurs in the long QT syndrome~\cite{Viskin1999,Viskin2021}, where the interval between waves Q and T of the electrocardiogram is prolonged. This interval correspond to the depolarization and repolarization of ventricular cells. A long QT means that the cells have longer plateaus and take more time to repolarize. If the potential spends more time in this range of values, it can prematurely activate depolarizing currents, causing EAD~\cite{Weiss2010}. Long QT often evolves to a type of polymorphic ventricular tachycardia known as \textit{torsades de pointes} (twisting of the points or peaks). Monomorphic ventricular tachycardias appear on the electrocardiogram as single amplitude oscillations and manifest as a stable spiral wave in the cardiac tissue, as in the case of a wave rotating a fixed, anatomical obstacle~\cite{Antzelevitch2001} (for example, a scar in the muscle). Polymorphic ventricular tachycardias, on the other hand, appear as oscillations with modulated amplitude and have a spiral with drifting or meandering core, i.e., the waves rotate around a functional core. Particularly, in \textit{torsades de pointes} the oscillations appear to twist around the electrocardiogram baseline, reversing its polarity~\cite{Viskin1999,Viskin2021}. Polymorphic ventricular tachycardias can deteriorate to ventricular fibrillation, leading to cardiac arrest and death. 

However, how the cellular manifestation (EAD) connects to the clinically observed tachycardias is not entirely clear~\cite{Weiss2010}. Typically, EAD and the resulting fluctuations at tissue level are irregular and seen as random phenomena. The experimental and computational work by Xie et al.~\cite{Xie2007} and Sato et al.~\cite{Sato2009} showed that the EAD oscillations are actually chaos and reentry can result from the loss of synchronization between patches of chaotic elements~\cite{Pecora1990,Heagy1995,Pecora1997}, evolving to complex spatiotemporal patterns of wandering spirals. Reentry was demonstrated in two- and three-dimensional simulations that represented the cardiac muscle, using detailed biophysical descriptions of cells~\cite{Beeler1977,Mahajan2008} (multidimensional conductance-based models) coupled as reaction-diffusion systems. Measures of a pseudoelectrocardiogram in these simulations results in rhythms that resemble polymorphic ventricular tachycardias and \textit{torsades de pointes}, thus unifying these phenomena.

In this work, we adapt and generalize their approach using a three-dimensional map-based membrane potential model, the logistic KTz~\cite{ktzlog2017}, capable of generating the behavior of cardiac cells and neurons by adjusting its six parameters. Though conductance-based models provide biophysically detailed simulations, models of cardiac cells are often described by dozens of differential equations with several adjustable parameters~\cite{Fenton2008b}. This creates problems such as difficult implementations and poor computational efficiency, specially in network and tissue simulations~\cite{ktzlog2017}. Therefore, obtaining simpler and more effective representations is an import part in the goal of understanding and preventing arrhythmias. The logistic KTz model was proposed to be a computationally efficient alternative for conductance-based models, retaining a wide range of biophysical behaviors with few equations and adjustable parameters.

Although proposed to simulate neuronal behaviors, it also presents the plateau spikes of cardiac cells. Pathological cardiac behaviors are also found, such as non-chaotic aperiodic spikes and chaotic EAD. We simulate the surface of the heart using square lattices with discrete diffusive coupling and nearest neighbors connections. Coupled map lattices are a good trade-off between the detailed biophysical models mentioned above and even simpler approaches, as cellular automata~\cite{Bub1998,Bub2005,Bub2013,Clayton2010,Christensen2015}. They are also specially suited for the study of spatiotemporal chaos~\cite{Kaneko2014}. Compared to conductance-based models, specially in network simulations, maps allow for an in-depth exploration of the parameter space because of the simpler mathematics and computational efficiency. However, literature on the topic is meager, with very few examples of map-based membrane potential models and coupled map lattices for cardiac simulations. In comparison, map-based models in neuroscience are far more common~\cite{Ibarz2011,review_mapas}.

Holden and Zhang~\cite{Holden1993a} use an unpublished three-dimensional map-based model with six parameters, proposed by Chialvo~\cite{Chialvo1995}, to study the effects of anisotropy in the propagation properties of the cardiac muscle simulated with a cubic lattice. The model appears to only vaguely reproduce the cardiac action potential. Rulkov~\cite{Rulkov2007} proposes a two-dimensional map for the cardiac action potential. The model makes a faithful representation, but requires several piecewise functions and up to ten parameters to achieve the desired behavior. Pavlov et al.~\cite{Pavlov2011} discretize the conductance-based Luo-Rudy model~\cite{Luo1991}, resulting in a four-dimensional model with eight parameters. The model has the advantage of inheriting the biological correspondence of the parameters, but requires several complicated polynomials to reproduce the properties of the ionic currents and the gating variables. It is also unclear if all these models can reproduce behaviors associated with pathologies, such as EAD.

To our best knowledge, this is all the literature available about the use of maps for cardiac simulations (with the exception of works that adopt a map-based approach for the study of alternans~\cite{Tolkacheva2003,Fenton2008}). Besides the richness of cardiac behaviors, computational efficiency and mathematical simplicity of the logistic KTz model, one major advantage is the ability to switch to neuronal behaviors only by adjusting its few parameters. Another goal in generalizing studies about the genesis of cardiac spirals is the modeling of spiral waves in the brain, a very little understood phenomenon the appears to be related to seizures and epilepsy~\cite{Winfree2003,Huang2004,Schiff2007,Huang2010,Viventi2011,Liu2018}. Although this case will not be dealt in this paper, we intend to use the insights gained here to model the neural case in a future work, taking advantage of the vast neuronal behaviors found in the KTz model.

Despite the simplicity of the KTz model, we generate a bifurcation diagram of the action potential duration (APD) similar to the biophysically detailed model and create an analogous simulations of reentry to the ones with reaction-diffusion models with conductance-based equations. We also calculate the Lyapunov exponent for the whole network and check the relation of chaos desynchronization with network size. \textit{Torsades de pointes} appear in a generic sampling of the network activity during spiral formation. Our results also indicate that high coupling between the cells can lead to bursting behavior occurring along with plateau spikes. We emphasize that such a complete assessment of cardiac dynamics has never been done before with map-based models. In addition to generalizing, our study extends and complements the results obtained previously with conductance-based models. The paper is organized as follows. In Sect.~\ref{sec:model}, we describe the logistic KTz model in Sect.~\ref{sec:ktz_log} and the network structure and coupling in Sect.~\ref{sec:net_struc}. In Sect.~\ref{sec:results}, we look at generating chaos in the single cell in Sect.~\ref{sec:single_cell} and the whole network in Sect.~\ref{sec:network}, then we check the desynchronization features in Sect.~\ref{sec:desync} and the spatiotemporal properties of the simulatons for a few examples in Sect.~\ref{sec:portrait}. We present our conclusions and ideas for future studies in Sect.~\ref{sec:conclusion}.

\section{Model}
\label{sec:model}

\subsection{Membrane potential}
\label{sec:ktz_log}

The logistic KTz model~\cite{ktzlog2017} describes the membrane potential of each cell by the three-dimensional map
\begin{align}
	x(t+1) &= f \left(\frac{ x(t) - Ky(t) + z(t) + I_{e} + I(t)}{T}\right), \label{x_ktz} \\
	y(t+1) &= x(t), \label{y_ktz} \\
	z(t+1) &= (1-\delta)z(t) - \lambda(x(t) - x_{r}), \label{z_ktz}
\end{align}
\noindent where $f$ is the logistic equation $f(u)=u/(1+|u|)$, $x(t)$ represents the membrane potential, $y(t)$ is a recovery variable and $z(t)$ is a slow adaptive current. $I_{e}$ is an external constant current and $I(t)$ is the the external time varying current (stimuli and/or "synapse"). Parameters $K$ and $T$ control the spiking dynamics. $\delta$ is the inverse recovery time of variable $z(t)$ and controls the refractory period. $\lambda$ controls the damping of the oscillations. The small values of $\delta$ and $\lambda$ makes $z(t)$ a slow variable. $x_{r}$ is a reversal potential, controlling bursting duration. Variables, time and parameters are all dimensionless. Time will be measured in time steps (ts). Cells are assumed to be points (dimensionless). The phase diagram of the model is drawn for parameters $x_{r}$ vs. $T$, with $K = 0.6$, $I_{e} = 0$, $\delta = \lambda = 0.001$ and no time varying current. An image of the diagram is reproduced in the next section (Fig.~\ref{fig:ISI_APD_diag_KTz_log}-(b)). Of particular interest is the region of the diagram between the cardiac and bursting phases. In this region, both behaviors mix chaotically and EAD oscillations are also found on the plateau.

\subsection{Network structure and coupling}
\label{sec:net_struc}
The usual approach to model the cardiac tissue starts with elementary equations from electromagnetic theory, such as Ohm's law and conservation of charges, applied to the intra- and extracellular domains, obtaining what is known as the bidomain model. Assuming the conductivity of both domains to be proportional, this model is simplified to the monodomain, which is a reaction-diffusion equation. If diffusion is constant in all directions, we obtain a diffusion tensor multiplied by the Laplacian of the potential. If we disregard anisotropy and assume that diffusion is the same in all directions, the tensor is replaced by a single constant and we obtain the common equation for general cardiac tissue models (see Clayton et al.~\cite{Clayton2010} for a proper deduction). This procedure is independent of the chosen membrane potential model and different conductance-based or differential equations models can be used to describe the transmembrane currents.

Because the heart muscle behaves macroscopically as a functional syncytium~\cite{Clayton2010}, i.e., cells closely connected to form a nearly continuous tissue, this approach approximates the tissue as a continuous space. However, close to the cellular scale, the heart is composed of discrete cells. The cardiac tissue is also anisotropic, with preferential propagation direction parallel to the muscle fibers and fewer connections in the transversal direction~\cite{Clayton2010}. It is also three-dimensional, but the thinner atrial muscle is reasonably approximated by two dimensions and some propagation properties in the ventricles can be studied two dimensions~\cite{Cherry2008,Alonso2016}. As we aim at a more general model, we retain the common assumptions of isotropy and two dimensions (for example, used in Xie et al.~\cite{Xie2007} and Sato et al.~\cite{Sato2009}), but we discretize space to fit with the map-based membrane potential model.

\begin{figure}
\centering
\resizebox{0.6\columnwidth}{!}{\includegraphics{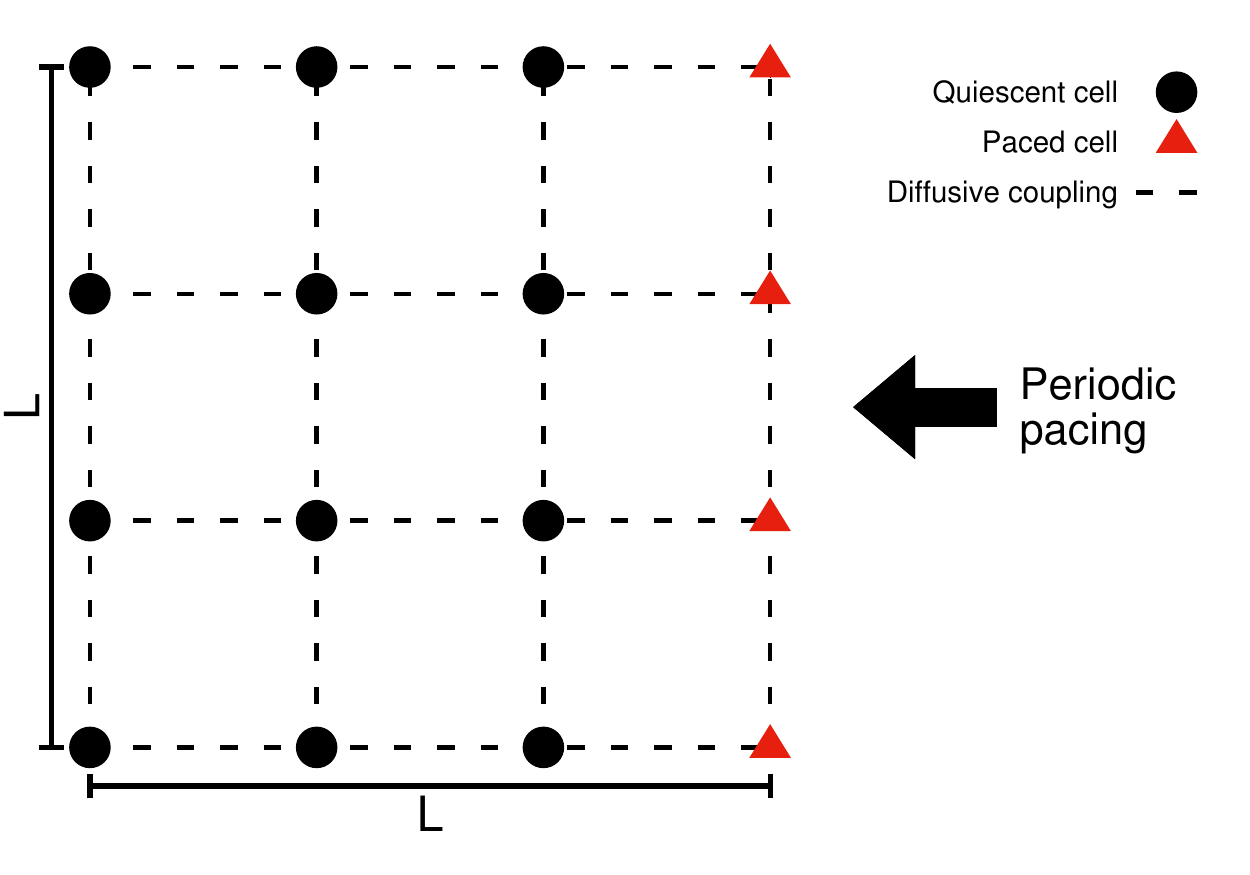}}
\caption{Network structure of the two dimensional lattice used in the simulations and the pacing protocol adopted, with $L = 4$.}
\label{fig:net_struc}
\end{figure}

This is achieved by discretizing the Laplacian in the aforementioned equation, using the second order central difference operator for the two spatial dimensions. This directly leads to a regular network, i.e., a lattice, with first nearest neighbors interactions (Von Neumann neighborhood) and discrete diffusive coupling, like in the coupled map lattices used previously for cardiac tissue~\cite{Holden1993a,Pavlov2011}. This type of coupling occurs in the gap junctions~\cite{Alonso2016} among heart cells, which are also found in the brain under the name of electrical synapses and allows for a fast and direct communication of the stimuli. Generically, the total diffusive current is given by
\begin{align}
	I_{i}(t) = \sum_{j}J_{ij}[x_{j}(t)-x_{i}(t)], \label{eq:I_eletr}
\end{align}
where $J_{ij}$ is the conductance of the junction (or diffusive constant), $i$ refers to the cell receiving ("postsynaptic" in the neuroscience jargon) and $j$ to the neighbor cell sending the stimulus ("presynaptic"). Aligned with the isotropy assumption, all couplings are of equal value ($J_{ij} = J$). Adopting open boundary conditions, we reach the network structure of the $N = L \times L$ lattice depicted in Fig.~\ref{fig:net_struc}, along with the periodic stimulation protocol used in the simulations. Quiescent and paced cells (rightmost column of the network) have the same parameters and, consequently, the same behavior. However, paced cells receive a constant intensity stimulus at $P$ intervals. The stimulus is a pulse, that is, longer than $1$ ts. $P$ is counted from the beginning of the pulse. All the paced cells in the rightmost column are paced simultaneously with the same stimulus. Paced cells can be thought of as pacemaker cells or located at the intersection with pacemaker cells of the heart's conduction system, like the sinoatrial node cells.

\section{Results}
\label{sec:results}

\subsection{Generating chaos in the single cell}
\label{sec:single_cell}

To simulate the arrhythmic tissue, we need to set the appropriate parameters. The first step is to define parameters that will generate the desired behavior in the cells composing the network. From the model description in Sect.~\ref{sec:net_struc}, we see that the network will also impose additional parameters: the coupling $J$ and the pacing period $P$. Both need to be chosen to generate spatiotemporal chaos in the network, which results from coupled chaotic oscillators. Therefore, the problem initially comes down to finding the KTz parameters and the pacing period that will generate chaos in the single cell, to later connect the cells and look for the appropriate coupling. Previous works followed a similar procedure of finding chaos in the cell, then assembling networks~\cite{Xie2007,Sato2009}.

\begin{figure}
\centering
\resizebox{1.0\columnwidth}{!}{\includegraphics{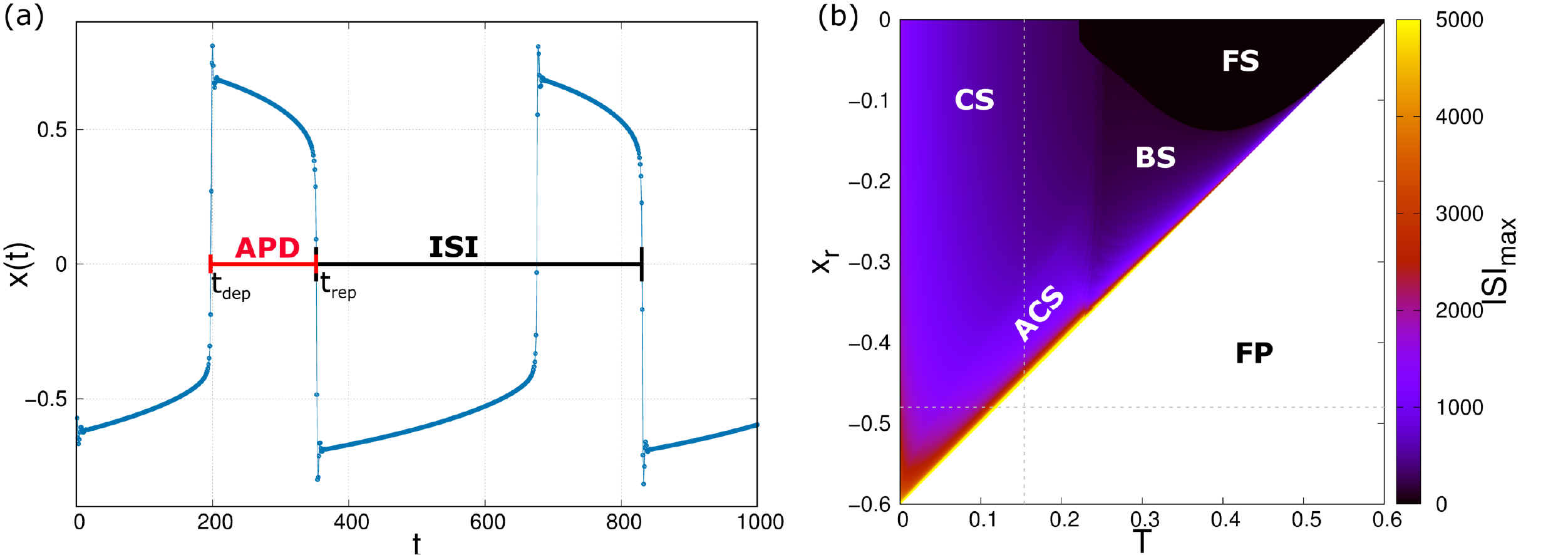}}
\caption{(a) Action potential duration (APD) and interspike interval (ISI) exemplified in an autonomous cardiac behavior ($T = 0.15$ and $x_{r} = -0.2$, the rest of the parameters follow the phase diagram in (b)) of the logistic KTz model. Initial conditions (IC) are $x(0) = y(0) = -0.5$ and $z(0) = 0$. (b) Phase diagram of the logistic KTz model ($K = 0.6$, $I_{e} = 0$ and $\delta = \lambda = 0.001$). The color scale represents the maximum ISI. Labels refers to the behaviors: cardiac spiking (CS), aperiodic cardiac spiking (ACS), bursting (BS), fast spiking (FS) and fixed point (FP). The intersection of the dashed lines ($T = 0.154$ and $x_{r} = -0.48$) indicates the FP we periodically stimulate to generate the excitable cardiac spikes in Fig.~\ref{fig:bifurc_APD_lyap_insets}. IC are $x(0) = y(0) = z(0) = 0$, the maximum number of iterations for each point is $t_{m} = 10^{5}$ ts with a discarded transient of $t_{t} = 5\times10^{4}$ ts and axis interval between points of $\Delta$$x_{r}$ = $\Delta$$T$ = $10^{-4}$. Colored figures online.}
\label{fig:ISI_APD_diag_KTz_log}
\end{figure}

To identify the aperiodic behavior that indicates chaos in the cell, we look at the action potential duration (APD). Since cardiac spikes take longer times than neuron spikes, APD is often the quantity of interest when assessing nonlinear properties of cardiac systems, while neuroscience applications often focus on the interspike interval (ISI)~\cite{Innocenti2007,Gu2013c,Xu2020,Guo2021}. Figure~\ref{fig:ISI_APD_diag_KTz_log}-(a) exemplifies the APD along with the ISI in a cardiac spike of the logistic KTz model. The APD is measured between the instants of depolarization ($t_{dep}$) and repolarization ($t_{rep}$), when the membrane potential $x(t)$ changes sign. Mathematically, the depolarization is identified when $x(t_{dep}+1)x(t_{dep}) < 0$ and $x(t_{dep}) < x(t_{dep}+1)$ and the repolarization when $x(t_{rep}+1)x(t_{rep}) < 0$ and $x(t_{rep}+1) < x(t_{rep})$. The i-th APD in a time series is defined as $APD_{i} = t_{i,rep} - t_{i,dep}$. Though we calculate the APD at $x = 0$, we note that it is usual in the literature to calculate at $90\%$ of repolarization~\cite{Alonso2016}.

We restrict our attention to the phase diagram of the logistic KTz model, shown in Fig.~\ref{fig:ISI_APD_diag_KTz_log}-(b), which plots the maximum ISI for each pair of parameters $T$ and $x_{r}$. Details about the phase diagram and the calculation of the ISI can be found in the original paper~\cite{ktzlog2017} and its Supplementary Information. To simulate the excitable heart muscle, the myocardial tissue, we need excitable behavior. We define excitable behavior as the spike answer of the resting state of a cell after an external stimulus. A way to generate this behavior is to adjust the parameters in the (negative) FP region of the phase diagram~\cite{Copelli2004}, near the autonomous phase of desired behavior. Dashed lines in Fig.~\ref{fig:ISI_APD_diag_KTz_log}-(b) indicate this point as $T = 0.154$ and $x_{r} = -0.48$, below the ACS phase.

\begin{figure}
\centering
\resizebox{1.0\columnwidth}{!}{\includegraphics{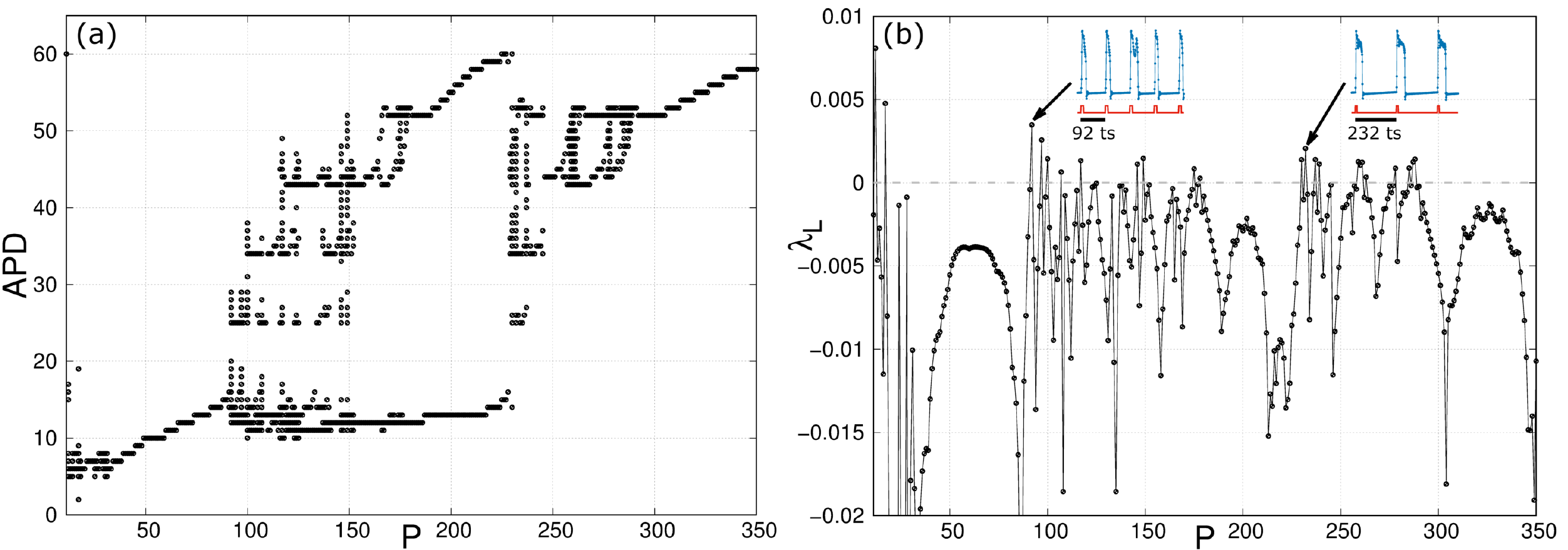}}
\caption{(a) Bifurcation diagram of the APD against the pacing period $P$ with excitable cardiac spiking ($T = 0.154$ and $x_{r} = -0.48$) and (b) the corresponding Lyapunov exponent. Single or a few repeating APD for a given pacing indicate periodic behavior, while several denote aperiodicity. Pacings with aperiodic behavior in (a) have positive exponent in (b), confirming chaos. A current pulse of intensity $I = 0.1$ and duration of $10$ ts was used. Insets in (b) show the membrane potential (blue) and current (red) for two values of the pacing: fast ($P = 92$ ts) and slow ($P = 232$ ts). In (a) and (b), the IC are $x(0) = y(0) = z(0) = 0$ and $t_{m} = 10^{6}$ ts. In (a), also $t_{t} = 9.5\times10^{4}$ ts. Colored figures online.}
\label{fig:bifurc_APD_lyap_insets}
\end{figure}

We periodically pace the cell with a current pulse with intensity $I = 0.1$ and duration of $10$ ts. Bifurcation diagrams of the APD plotted against the pacing period $P$ (measured from the start of each stimulus) allow us to systematically see the behavior of the time series for each pacing. If the cell response is periodic for a given pacing $P$, we will observe a single or a few repeating APD values in the diagram. If the cell response is aperiodic, infinitely many APD values may appear in a given range for that pacing. However, it is worth noting that the APD are distributed in discrete values, given the discrete-time nature of the model (as  opposed to differential equations, which generate a continuum of values in aperiodicity). Figure~\ref{fig:bifurc_APD_lyap_insets}-(a) shows the bifurcation diagram for the aforementioned point. The shape of the diagram strongly resembles the shape of one of the bifurcation diagrams obtained by Sato et al.~\cite{Sato2009} in their biologically plausible model (Fig.S3 of their Supporting Information), resembling the diagram made from experiments on a ventricular cell of the rabbit heart (Fig.S1 of their Supporting Information). In the logistic KTz, bifurcation diagrams with this shape were mostly found below the ACS phase of the phase diagram, with parameter $T$ varying between around $0.13$ and $0.2$.

To confirm that the aperiodicity is chaos, we calculate the Lyapunov spectrum for the single cell, defined as
\begin{align}
	\lambda_{l} &= \lim_{t_{m} \rightarrow \infty} \frac{1}{t_{m}} ln \left|\Lambda_{l}\right| \qquad (l = 1,2,3), \label{eq:lyap}	
\end{align}
where $\Lambda_{l}$ are the eigenvalues of the matrix resulting from the product of the Jacobians after $t_{m}$ iterations. The $3 \times 3$ Jacobian at time $t$ is
\begin{align}
	\textbf{\textit{J}}(t) &= \frac{\partial(x(t+1),y(t+1),z(t+1))}{\partial(x(t),y(t),z(t))} \label{eq:jacob_single} \\
	&=
	\begin{pmatrix}
			\frac{1}{T[1+ \left|u(t)\right|]^{2}} & \frac{-K}{T[1+ \left|u(t)\right|]^{2}} & \frac{1}{T[1+ \left|u(t)\right|]^{2}} \\
			1 & 0 & 0\\
			-\lambda & 0 & 1-\delta\\
	\end{pmatrix}
	\label{eq:jacob_matrix_single}
	,
\end{align}
where $u(t) = [x(t) - Ky(t) + z(t) + I_{e} + I(t)]/T$ is the argument of the logistic function in Eq.~\ref{x_ktz}. The Lyapunov spectrum is approximated using the Eckmann-Ruelle method~\cite{eckmann1985}, which consists in successively triangularize the matrices using a LU decomposition during the iterations. Figure~\ref{fig:bifurc_APD_lyap_insets}-(b) shows the largest Lyapunov exponent $\lambda_{L}$ of the mapping. Positive values of the exponent correspond to aperiodic behaviors in Fig.~\ref{fig:bifurc_APD_lyap_insets}-(a). The insets show the chaotic time series for fast ($P = 92$ ts) and slow ($P = 232$ ts) pacing.

\subsection{Generating chaos in the network}
\label{sec:network}

We couple the logistic KTz elements using the network structure of Fig.~\ref{fig:net_struc} and the discrete diffusive coupling of Eq.~\ref{eq:I_eletr}. The network is homogeneous, with all the cells having the same parameters as defined in the previous section ($T = 0.154$ and $x_{r} = -0.48$), including quiescent and paced cells. To simplify the notation, we take a numbering scheme for the cells as $i = 1,...,N$, from left to right and bottom to top. In the square lattice, the diffusive current of Eq.~\ref{eq:I_eletr} for cell $i$ reads $I_{i}(t) = J[x_{i+1}(t) + x_{i-1}(t) + x_{i+L}(t) + x_{i-L}(t) - 4x_{i}(t)]$. Because of the open boundary conditions, the cells at the rightmost column have $x_{i+1}(t) = 0$, cells at leftmost column have $x_{i-1}(t) = 0$, cells at top row have $x_{i+L}(t) = 0$ and cells at bottom row have $x_{i-L}(t) = 0$.

The initial conditions are assumed $x_{i}(0) = y_{i}(0) = -0.5$ and $z_{i}(0) = 0$ to resemble the resting state. To generate activity in the network, we use the same type of pacing of the single cell, but now delivered to all the cells at the rightmost column. The expected response of the healthy tissue is to simply generate periodic plane wavefronts that propagates from right to left. We select the value $P = 92$ ts for the pacing, as it has the highest value of the Lyapunov exponent ($\lambda_{L} \approx 0.0035$) for the single cell (Fig.~\ref{fig:bifurc_APD_lyap_insets}-(b)), with pacing above the longest APD ($APD < P$). To create an initial heterogeneity, a small difference on the IC is given to the paced column as $x_{i}(0) = -0.5 + \epsilon_{x,i}$ and $y_{i}(0) = -0.5 + \epsilon_{y,i}$, where each $\epsilon$ is a pseudorandom number with uniform distribution between $\pm5\times10^{-8}$ and cell $i$ belongs to the rightmost column.

\begin{figure}
\centering
\resizebox{1.0\columnwidth}{!}{\includegraphics{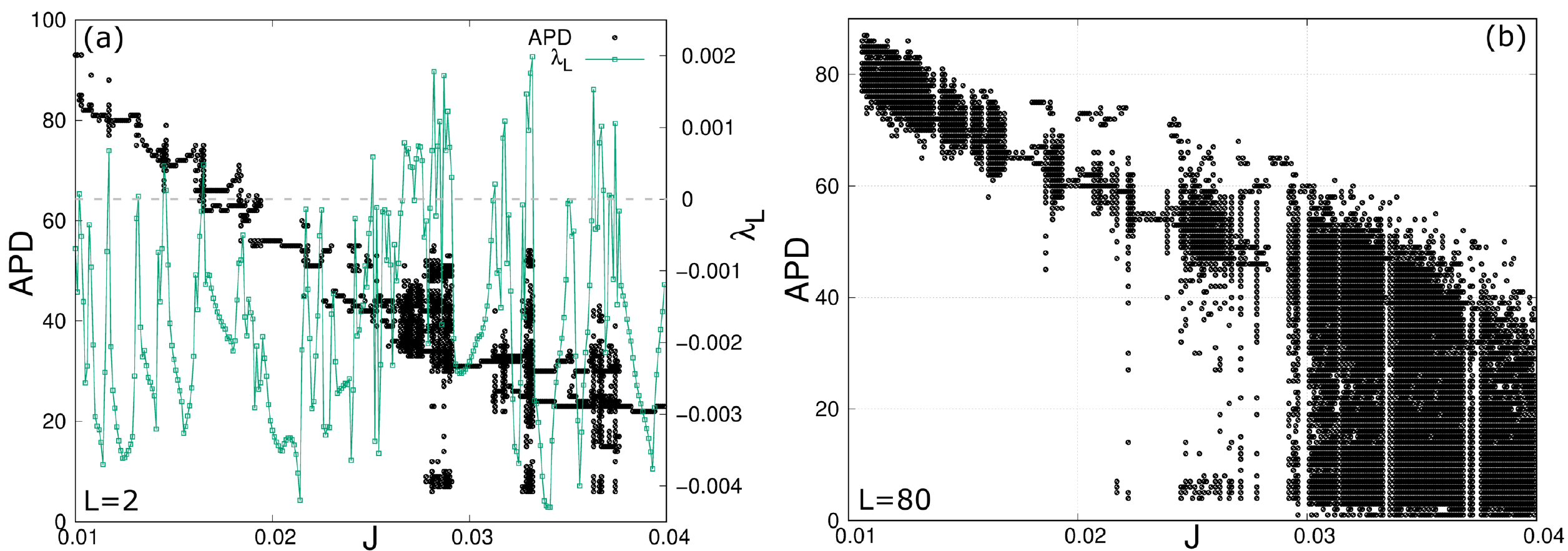}}
\caption{Bifurcation diagram of the APD against the coupling $J$ for network sizes (a) $L = 2$ and (b) $L = 80$ (black points). Pacing is $P = 92$ ts and applied in the rightmost column. In (a), the bifurcation was measured at the bottom left vertex cell ($i = 1$). The largest Lyapunov exponent $\lambda_{L}$ from the network spectrum is also plotted in green (right axis), showing that aperiodic regions correspond to chaos. In (b), the bifurcation was measured in one of the central cells of the network ($i = \frac{N-L}{2} = 3160$). IC for the network are specified in the main text. For APD (in (a) and (b)) and $\lambda_{L}$, $t_{m} = 10^{5}$ ts and $\Delta$$J$ = $10^{-4}$. APD also have $t_{t} = 9\times10^{4}$ ts and 10 realizations for each point. Colored figures online.}
\label{fig:bifurc_exp_L=2_L=80}
\end{figure}

One more constant is needed to be defined: the diffusion coefficient $J$ (not to be confused with the Jacobian matrix $\textbf{\textit{J}}$). The coupling can have dramatic effects on the dynamics of the cells, both theoretically and experimentally~\cite{Clayton2010}. This constant also affects the capacity of chaotic oscillators to synchronize~\cite{Pecora1997}. Hence, we select one of the cells of the network and again look for aperiodicity in the APD duration, but now for different values of $J$. However, in doing so, we are inferring the network behavior based on a single network cell. Turns out that this bifurcation diagram is a very good chaos indicator for the whole network. To check this statement, we generalize the calculation of the Lyapunov exponent for the entire network.

As the network is composed of $N$ cells, each one described by Eqs.~\ref{x_ktz}-\ref{z_ktz}, the Jacobian will be a $3N \times 3N$ matrix and the resulting Lyapunov spectrum of Eq.~\ref{eq:lyap} will now have $3N$ exponents ($l = 1,...,3N$). For clarity, we adopt the vector notation for the variables of cell $i$ as $\vec{x}_{i}(t) = (x_{i}(t),y_{i}(t),z_{i}(t))$. The network Jacobian is defined as
\begin{align}
	\textbf{\textit{J}}(t) &= \frac{\partial(\vec{x}_{1}(t+1),...,\vec{x}_{N}(t+1))}{\partial(\vec{x}_{1}(t),...,\vec{x}_{N}(t))} \label{eq:jacob_net} \\
	&=
	\begin{pmatrix}
			\textbf{\textit{J}}_{1,1}(t) & \cdots & \textbf{\textit{J}}_{1,N}(t) \\
			\vdots & \ddots & \vdots \\
			\textbf{\textit{J}}_{N,1}(t) & \cdots & \textbf{\textit{J}}_{N,N}(t) \\
	\end{pmatrix}
	\label{eq:jacob_matrix_net}
	,
\end{align}
where submatrices $\textbf{\textit{J}}_{i,j}$ contain the derivatives between variables of cells $i$ and $j$. Given the regular topology with nearest neighbors interactions, $\textbf{\textit{J}}$ will be a sparse matrix, with most submatrices zero given the lack of dependence between the variables of the cells.

The submatrices with non-zero terms will be of two types: the ones with derivatives between the variables of cell $i$ and the ones with derivatives between variables of neighboring cells $i$ and $j$. Each one follows a generic format, irrespective of the cells involved. The first one is
\begin{align}
	\textbf{\textit{J}}_{i,i}(t) =
	\begin{pmatrix}
			\frac{1-n_{i}J}{T[1+ \left|u_{i}(t)\right|]^{2}} & \frac{-K}{T[1+ \left|u_{i}(t)\right|]^{2}} & \frac{1}{T[1+ \left|u_{i}(t)\right|]^{2}} \\
			1 & 0 & 0 \\
			-\lambda & 0 & 1-\delta \\
	\end{pmatrix}
	\label{eq:subjacob_ii}
	,
\end{align}
almost identical to Eq.~\ref{eq:jacob_matrix_single}, except for the first term, whose derivative is modified because of the dependence of Eq.~\ref{eq:I_eletr} with $x_{i}(t)$. Likewise, $u_{i}(t)$ is the argument of the logistic function and $n_{i}$ is the coordination number (i.e. number of "synapses") for cell $i$. The second is
\begin{align}
	\textbf{\textit{J}}_{i,j}(t) =
	\begin{pmatrix}
			\frac{J}{T[1+ \left|u_{i}(t)\right|]^{2}} & \phantom{000}0\phantom{000} & \phantom{000}0\phantom{000} \\
			0 & \phantom{000}0\phantom{000} & \phantom{000}0\phantom{000} \\
			0 & \phantom{000}0\phantom{000} & \phantom{000}0\phantom{000} \\
	\end{pmatrix}
	\label{eq:subjacob_ij}
	,
\end{align}
with the first term being the only non-zero because of the dependence of Eq.~\ref{eq:I_eletr} with $x_{j}(t)$. After assembling the full Jacobian with the submatrices, we proceed as in the last section, calculating the Lyapunov spectrum using the Eckmann-Ruelle method.

Figure~\ref{fig:bifurc_exp_L=2_L=80}-(a) shows the largest Lyapunov exponent obtained for the minimum size network $L = 2$ against coupling $J$. Concurrently, the APD bifurcation diagram, measured at the bottom left corner cell of the network. This indicates that aperiodicity measured in the cell corresponds to chaos in the network and that a proper value of $J$ is needed for a chaotic response to occur. Figure~\ref{fig:bifurc_exp_L=2_L=80}-(b) shows the APD bifurcation diagram measured in one of the central cells for $L = 80$, our target network size in this study. In this case, the calculation of the Lyapunov spectrum is very computationally demanding and we were not able to obtain the result in time, even with several efficiency tweaks to our code.

We can observe a general downward trend for the maximum APD with the increase of $J$. The diagram in Fig.~\ref{fig:bifurc_exp_L=2_L=80}-(b) can be roughly divided into three main regions. Below $J = 0.02$, the coupling is too weak to generate interesting activity and wavelets fails to spread transversely when synchronization is lost. Above $J = 0.03$, the cardiac (plateau) spikes start being replaced by bursting. They appear as strong oscillations at the beginning or end of the plateau or completely replace the cardiac spikes. This will be illustrated in Sect.~\ref{sec:portrait}, along with the complex spatiotemporal activity that results. The intermediate $J$ values usually yield the best results for our objective of simulating cardiac arrhythmias. Transverse propagation of the broken plane waves leads to reentry and APD is predominantly plateau spikes (with EAD).

\subsection{Desynchronization of chaos}
\label{sec:desync}

Regions of aperiodic behavior in the APD bifurcation diagram indicate the occurrence of synchronized or desynchronized chaos. Conversely, regions with a single or few APD indicate (synchronized) periodic behavior. As we are dealing with a chaotic system composed of identical cells with differences in the IC, chaos will amplify the small differences in the diffusive coupling of Eq.~\ref{eq:I_eletr} and lead to desynchronization, provided that a certain "critical" size of the network is reached~\cite{Heagy1995}. Although there are varied approaches to quantify the collective behavior and synchronization in similar systems~\cite{Mainieri2005,Qin2014,Naze2015,Yao2019,Zhang2020}, we adapt the basic approach of Xie et al.~\cite{Xie2007}. But, instead of just averaging the standard deviations of membrane potentials calculated at each time step, we modify the expression to use the column-based symmetry of stimulus propagation.

We calculate the standard deviation of the membrane potentials $x_{(r,c)}(t)$ for each column in time $t$ and then average these values. This mean standard deviation is then averaged in time. Indexes $r$ and $c$ represent rows and columns, respectively. The following equation is used:
\begin{align}
	\left\langle\overline{\sigma}_{x}\right\rangle &= \lim_{t_{m} \rightarrow \infty} \frac{1}{t_{m}-t_{t}}\sum_{t=t_{t}}^{t_{m}}\left(\frac{1}{L}\sum_{c=1}^{L}\sqrt{\frac{1}{L}\sum_{r=1}^{L}x_{(r,c)}^{2}(t)-\left[\frac{1}{L}\sum_{r=1}^{L}x_{(r,c)}(t)\right]^{2}}\right) \label{eq:sigma1}	\\
				&= \lim_{t_{m} \rightarrow \infty} \frac{1}{t_{m}-t_{t}}\sum_{t=t_{t}}^{t_{m}}\overline{\sigma}_{x}(t), \label{eq:sigma2}	
\end{align}
where $t_{m}$ is the maximum number of iterations and $t_{t}$ is the transient discarded. This expression relies on the columnar symmetry of the wavefronts generated by the stimuli, increasing its value as the cells in each column start firing out of sync.

\begin{figure}
\centering
\resizebox{0.8\columnwidth}{!}{\includegraphics{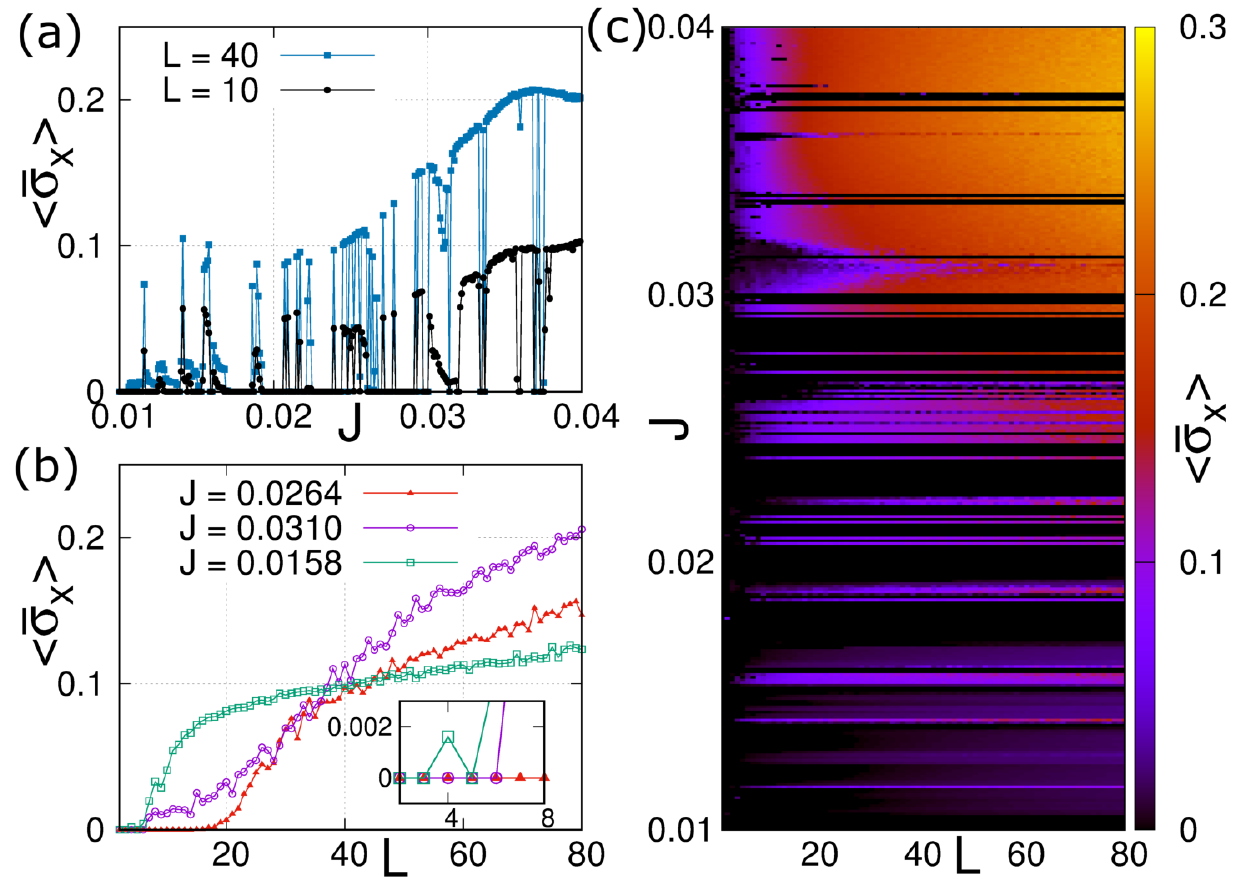}}
\caption{Standard deviation $\left\langle\overline{\sigma}_{x}\right\rangle$ of the membrane potential in each column, averaged among columns and time: (a) plot against coupling $J$ and (b) network size $L$. Desynchronization in (a) is in good agreement with aperiodic regions in Fig.~\ref{fig:bifurc_exp_L=2_L=80}-(b). Inset in (b) to clarify some of the critical values of desynchronization. (c) represents $\left\langle\overline{\sigma}_{x}\right\rangle$ in the color scale, varying $J$ and $L$. It is visible that most $J$ values desynchronize for small network sizes. IC for the network are described in the main text. In (a)-(c), $t_{m} = 10^{5}$ ts, $t_{t} = 5\times10^{4}$ ts and $\Delta$$J$ = $10^{-4}$. 100 realizations for each point in (a) and (b). 10 realizations in (c). Colored figures online.}
\label{fig:sigma_x_inset}
\end{figure} 

Figure~\ref{fig:sigma_x_inset}-(a) shows $\left\langle\overline{\sigma}_{x}\right\rangle$ against $J$ for two network sizes. Synchronized/desynchronized regions seem to agree with periodic/aperiodic regions of Fig.~\ref{fig:bifurc_exp_L=2_L=80}-(b). Counterintuitively though, increasing the coupling in this range increases desynchronization. Synchronization is likely to increase as we increase the coupling above this range. Figure~\ref{fig:sigma_x_inset}-(b) shows $\left\langle\overline{\sigma}_{x}\right\rangle$ against the network size $L$ for three different couplings. $J = 0.0264$ desynchronizes in $L = 15$ and $J = 0.0310$ in $L = 7$. $J = 0.0158$ shows a small increase in $\left\langle\overline{\sigma}_{x}\right\rangle$ for $L = 4$, followed by zero in $L = 5$. This is because desynchronization starts close to $t_{m}$ in $L = 4$ and above this value in $L = 5$. Figure~\ref{fig:sigma_x_inset}-(c) combines (a) and (b) in a diagram, showing that most couplings lose synchronization for small system sizes and indicates that chaotic regions are the same for sizes above $L = 40$ (at least up to $L = 80$). Notice that $L = 2$ is synchronized for all $J$, despite chaos for some values (Fig.~\ref{fig:bifurc_exp_L=2_L=80}-(a)).  

\subsection{Spatiotemporal portrait of reentry}
\label{sec:portrait}

\begin{figure}
\centering
\resizebox{1.0\columnwidth}{!}{\includegraphics{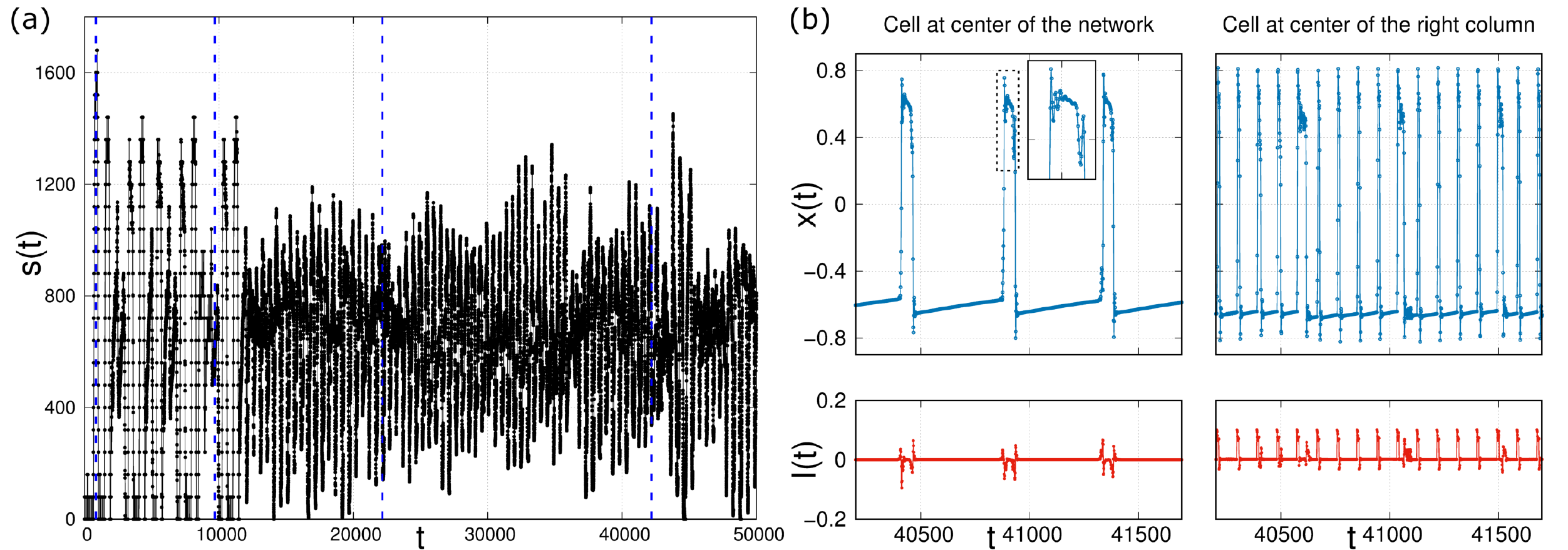}}
\caption{(a) Activity of the network, showing the evolution from plane waves to reentry and spiral waves with $L = 80$, $P = 92$ ts (applied to all cells at rightmost column) and $J = 0.0257$. Dashed lines correspond to panels $t = 840$ ts, $t = 9700$ ts, $t = 22170$ ts and $t = 42200$ ts in Fig.~\ref{fig:net_card}. (b) Membrane potential and current of the cell at the center of the network and center of the stimulated column. EAD oscillation is seen in the inset. IC for the network are described in the main text. Colored figures online.}
\label{fig:activity_potential_card}
\end{figure}

\begin{figure}
\centering
\resizebox{1.0\columnwidth}{!}{\includegraphics{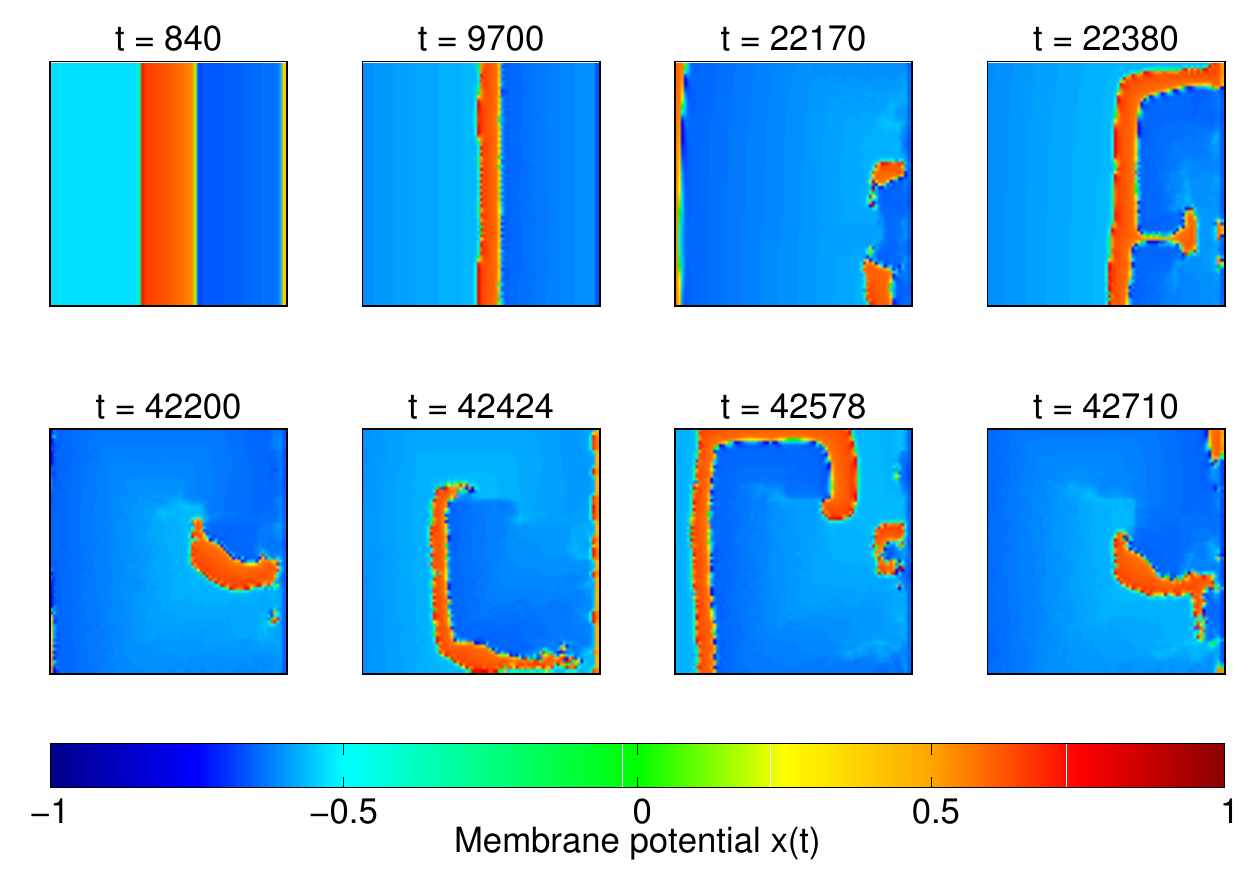}}
\caption{Spatiotemporal portrait of the network with $L = 80$, $P = 92$ ts (applied to all cells at rightmost column) and $J = 0.0257$. Plane waves ($t = 840$ ts), desynchronized waves without transverse propagation ($t = 9700$ ts), reentry ($t = 22170$ ts and $t = 22380$ ts) and developed spiral wave ($t = 42200$ ts to $t = 42710$ ts) are seen (videos are available online). The color scale represents the membrane potential. Colored figures online.}
\label{fig:net_card}
\end{figure} 

To demonstrate reentry and spiral wave formation, we select the value $J = 0.0257$. Figure~\ref{fig:activity_potential_card}-(a) shows the activity $s(t)$, measured as the number of cells with $x_{i}(t) > 0$ in each instant $t$, i.e., $s(t) = \sum_{i}^{N} \Theta(x_{i}(t))$, where $\Theta(x_{i}(t))$ is the Heaviside step function. The spatiotemporal portrait is in panels of Fig.~\ref{fig:net_card}. Dashed lines in Fig.~\ref{fig:activity_potential_card}-(a) indicate the instants that will be described in the next paragraphs (videos are also available online). Figure~\ref{fig:activity_potential_card}-(b) shows the membrane potential and current of a central cell of the network and central cell of the stimulated column. EAD oscillations are seen at the end of the plateau of most network cells (inset). The firing frequency of the paced cell is evidently higher, showing stimulation failure, with the network not propagating the stimulus each time the paced area is stimulated. 

Initially, plane (synchronized) wavefronts propagate from right to left in $t = 840$ (Video 1), with the activity evolving orderly with increases and decreases of the size of a column. Disorder is perceived in the activity with increments or decrements of different sizes than $L$. Initially, wavefronts slightly desynchronize, but without generating transverse propagation, as in $t = 9700$ ts (Video 2). The network fully desynchronizes around $t = 11500$ ts, with a marked difference in the activity pattern. Panels $t = 22170$ ts and $t = 22380$ ts show the reentry, with the stimulus propagating backwards (Video 3). Refractory regions form conduction blocks. The wavelets circulate their refractory trail (darker blue), entering regions recovered to the resting state (lighter blue). As this process evolves, the tissue develops spatiotemporal complexity, with the occurrence of spiral waves between $t \approx 33000$ ts and $t \approx 44600$ ts.

Because of stimulation failure and chaos desynchronization occurring in the pacing column or nearby, spirals form close to there. Panels $t = 42200$ ts to $t = 42710$ ts show a complete cycle of a spiral (Video 4). Comparing panels $t = 42200$ ts and $t = 42424$ ts, we can see how the refractory region and resting state alternate from one figure to the other. This simulation strongly resembles the results obtained by Xie et al.~\cite{Xie2007} in a conductance-based model. Noting that neuronal spikes in the logistic KTz range between $5$ and $10$ ts and real neurons between $1$ and $2$ ms, we can define a conversion factor of $1.5/7.5 = 0.2$ ms/ts. For the spiral wave in this example, we have a period of $102$ ms and frequency of $9.8$ Hz, close to experimental values of $\approx 100$ ms and $\approx 10$ Hz~\cite{Winfree2003}. Though the KTz model generically reproduces the cardiac spike, this factor also highlights that the plateaus in the phase diagram are of short duration. Cardiac action potentials of short duration are found in nature, such as in the mouse heart~\cite{Winfree2003,Bondarenko2017}.

\begin{figure}
\centering
\resizebox{0.6\columnwidth}{!}{\includegraphics{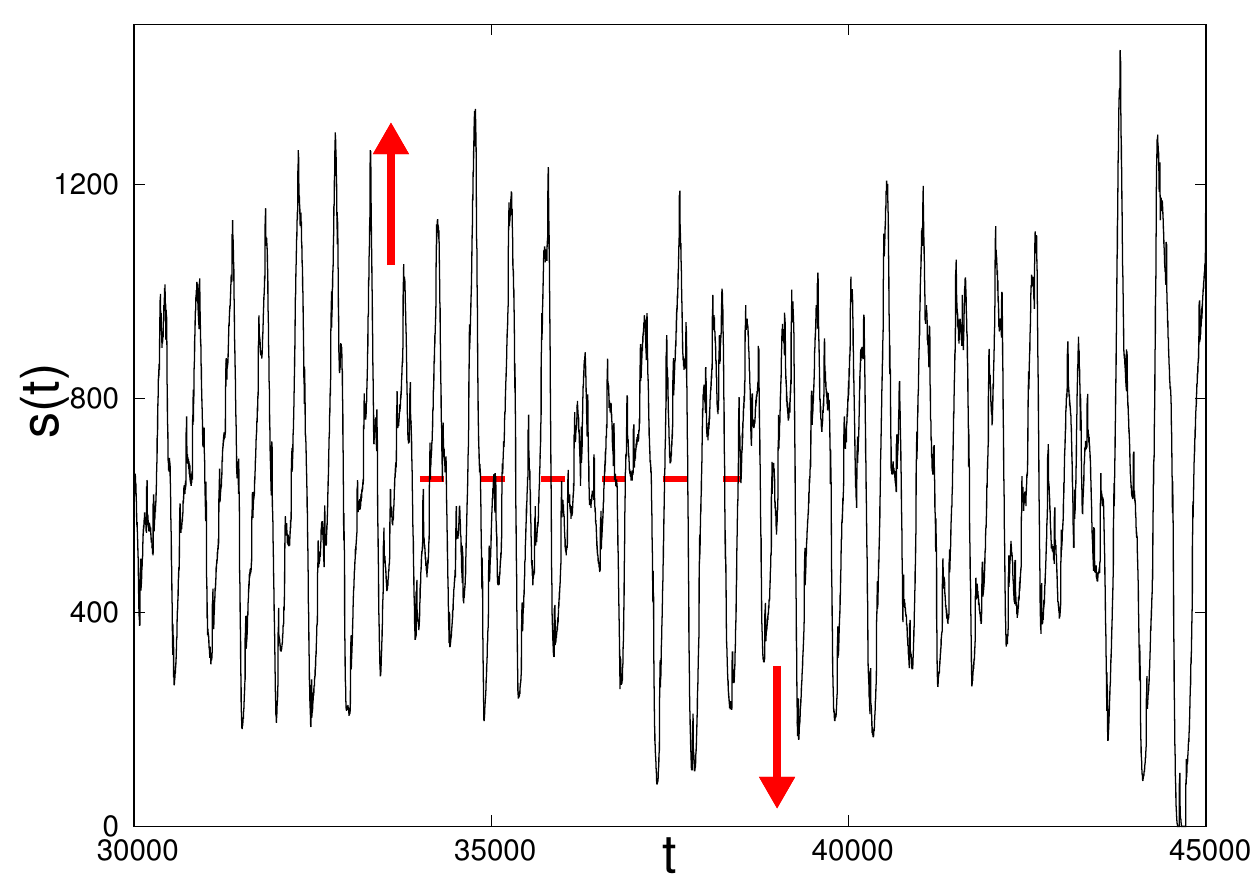}}
\caption{\textit{Torsades de pointes} (twisting of the points or peaks) is visible in the activity during spiral formation. The direction of the peaks are indicated by the arrows before and after the twisting. The dashed line represents an imaginary baseline of the electrocardiogram at $t = 650$ ts. Same time series from Fig.~\ref{fig:activity_potential_card}-(a).}
\label{fig:torsade}
\end{figure} 

Inspecting the activity more closely during the interval of spiral formation reveals a striking result: the oscillations strongly resembles \textit{torsades de pointes}, as can be seen in Fig.~\ref{fig:torsade}. For a comparison, readers are referred to the figure in the original article by Dessertenne, who first described \textit{torsades de pointes} in 1966, reprinted in Ref.~\cite{Dessertenne1990}. Not only shows the modulated oscillations similar to polymorphic ventricular tachycardia, but also the twisting of the peaks around an imaginary baseline, with the peaks in the example initially pointing upwards and, after the twist, pointing downwards. This result is surprising given the fact that we sampled the behavior of the whole network using a completely generic measure as the number of cells firing at an instant. Nevertheless, this result would be expected in a simulated electrocardiogram during these moments.

\begin{figure}
\centering
\resizebox{1.0\columnwidth}{!}{\includegraphics{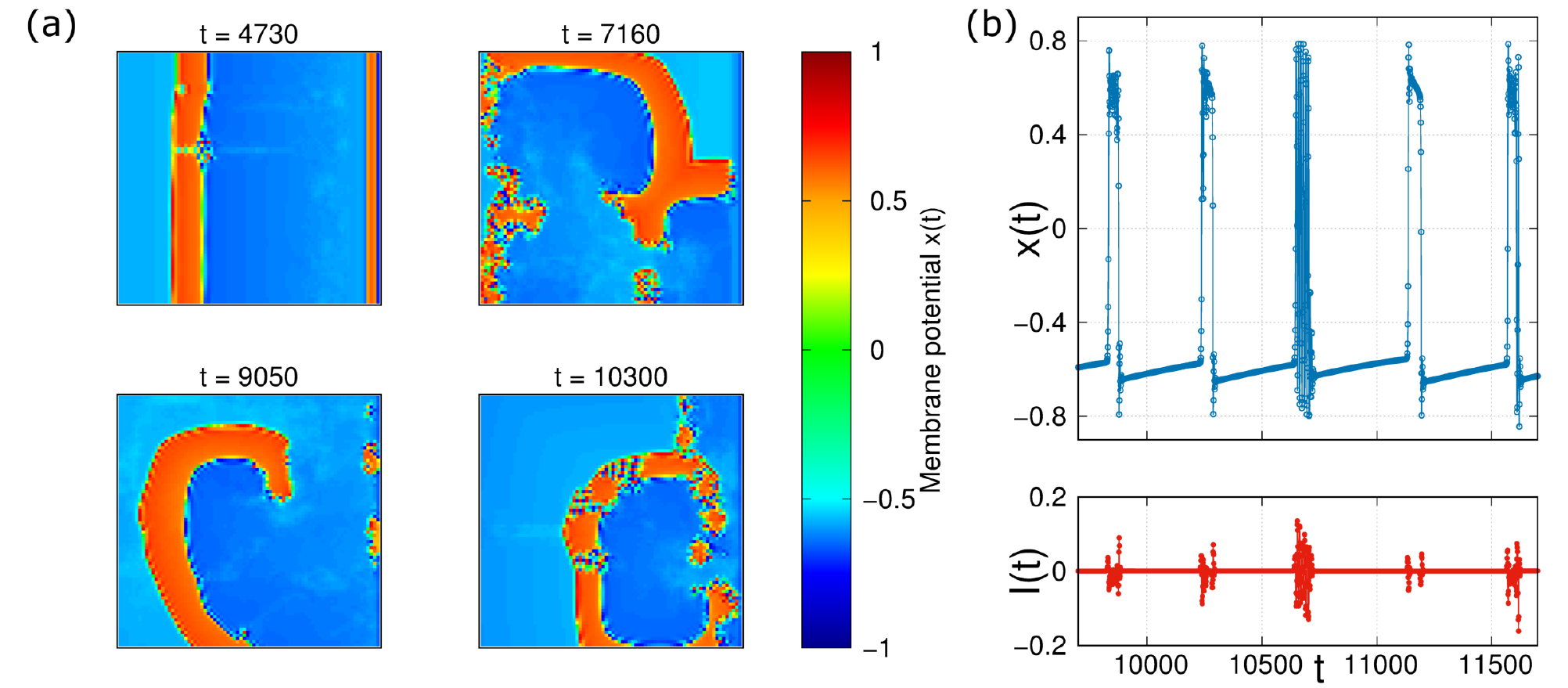}}
\caption{(a) Stronger coupling leads to bursting with $L = 80$, $P = 92$ ts (applied to all cells at rightmost column) and $J = 0.031$. Wavefronts are composed of synchronized cardiac spiking and bursting. The color scale represents the membrane potential. (b) Membrane potential and current of one of the central cells of the network, showing plateau spikes, bursting and plateaus with strong oscillations. IC for the network are specified in main text. Colored figures online.}
\label{fig:rede_x_I_card}
\end{figure} 

Figure~\ref{fig:rede_x_I_card}-(a) shows the spatiotemporal portrait for the strong coupling value $J = 0.031$ (Video 5). Figure~\ref{fig:rede_x_I_card}-(b) shows the membrane potential and current of one of the central cells of the network, displaying bursting along with cardiac spikes. Cardiac spikes also frequently show strong oscillations in the plateau. This is caused by a sensibility of the cells to high amplitude oscillations in the current, visible in the plot. Though plateau spikes still predominate, this reflects in the spatiotemporal portrait as wavefronts formed by regions with cardiac spiking along regions of bursting, all synchronized to compose the wavefront. This is specially evident in panel $t = 10300$ ts. The presence of bursting also increases the spatiotemporal complexity in the tissue, leading to reentry and spirals. Multiple spirals are seen during some instants.

\section{Conclusion}
\label{sec:conclusion}

We applied the map-based model, the logistic KTz, to simulate reentry and arrhythmia using chaos desynchronization in a two-dimensional lattice. First, we searched the phase diagram of the model for parameters that generated a bifurcation diagram of the APD against the period of pacing $P$ which contained aperiodicity. This bifurcation diagram resembled the one made from a biophysically detailed model, which in turn is similar to the diagram made from experiments on a cardiac cell (Figs. S3 and S1, respectively, from the Supporting Information of Sato et al.~\cite{Sato2009}). We confirmed that the aperiodic regions are chaos by calculating the Lyapunov exponent. Then we proceed to simulate the cardiac tissue.

Using these parameters for all cells and discrete diffusive coupling, we assembled a $L \times L$ regular network with nearest neighbors interactions. All the cells in the rightmost column where paced with the period $P$ that presented one of the highest values of the Lyapunov exponent for the single cell case. Initial conditions where chosen to resemble the resting state, with a small difference given to the paced cells. To define the last parameter required, the coupling constant $J$, we generated the bifurcation diagram of the APD against $J$ for a selected cell of the network. To confirm that the aperiodic values in the diagram corresponded to chaos in the network, we generalized the Lyapunov exponent calculation for the entire network. We demonstrated the correspondence of aperiodicity in the APD diagram and positive Lyapunov exponent for a small network, but we were unable to calculate this value for larger networks. For the larger network, we defined $J$ using the bifurcation diagram of the APD made in one the central cells.

Calculating an average standard deviation of the membrane potential of columns, we analyzed the chaos desynchronization properties of the network for different couplings $J$ and system sizes $L$. Overall, desynchronization increased with $J$ in the considered range and most values desynchronized for small sizes. This range could be roughly split into three regions: small, intermediate and high coupling. With small coupling, desynchronized waves failed to spread and generate reentry. High coupling led to high amplitude oscillations in the diffusive current, which turned some of the cardiac spikes into bursts of spikes without plateau or mixed both behaviors. Intermediate values generated the best results for the purpose of simulating reentry. Snapshots of the network showed the spatiotemporal portrait, with desynchronization of plane waves, reentry and the formation of a spiral wave. In general, cells exhibited a single EAD-like oscillations in the plateau. Surprisingly, the generic measure of the activity during spirals strongly resembled polymorphic ventricular tachycardia and \textit{torsades de pointes}, commonly seen in the electrocardiogram. We also showed snapshots of the network with high coupling, exhibiting reentry and spiral waves, but with wavefronts composed of uncommon substructures of synchronized plateau spikes and bursting.

There is some literature about cardiac bursting in models~\cite{Tong2005,Bondarenko2017} and experiments with spiral waves~\cite{Bub1998,Bub2005,Bub2013}. Though not clear in all those studies, such "bursts" are very likely bursts of spikes with plateau, even if the plateau duration is small or reduced. Bondarenko and Shilnikov~\cite{Bondarenko2017} studied a mouse ventricular model that presented bursts of spikes with very short duration, considering that the normal cardiac APD is already short in this animal. Short interval between stimuli can also shorten the cardiac plateau and refractory period~\cite{Winfree2003,Clayton2010,Alonso2016}. But the burst of Fig.\ref{fig:rede_x_I_card}-(b) is of spikes with the same APD of neurons in the logistic KTz model. The fifth action potential in Fig.\ref{fig:rede_x_I_card}-(b) begins with an oscillating plateau (EAD) and ends with an additional spike without plateau. Though not pictured in Fig.\ref{fig:rede_x_I_card}-(b), multiple additional spikes can occur at the end of the plateau. This behavior is also present in the transition between the CS and BS in the phase diagram of Fig.\ref{fig:ISI_APD_diag_KTz_log}-(b) (see Fig. 7-L from the paper that proposed the logistic KTz~\cite{ktzlog2017}). Rabbit ventricular cells with EAD exhibited something similar (see Fig. 6 of Yan et al.~\cite{Yan2001}), yet the spikes at the end of the action potential still have some longer APD than neurons. Given the simplified nature of the KTz, this is likely how this experimental behavior is represented in the model.

However, the wavefronts composed of synchronized bursts and plateaus are a novel result, yet of uncertain biophysical relevance. Himel et al.~\cite{Himel2013} showed experimentally that weak coupling in cardiac cells with EAD creates more pronounced oscillations in the plateau. Though not depicted in the results, the region with weak coupling (below $J < 0.02$) in Fig.~\ref{fig:bifurc_exp_L=2_L=80}-(b) generated multiple oscillations in the plateau, in contrast with a single oscillation (inset of Fig.~\ref{fig:activity_potential_card}-(b)) found with intermediate couplings ($J$ between $0.02$ and $0.03$). This suggest that, if possible, experimentally increasing the coupling could lead to burst-like spikes. Jiang and Hou~\cite{Jiang2015} demonstrated the formation of substructures in the arms of a stable spiral simulated with the Hindmarsh-Rose (HR) model. In a recent paper~\cite{ktz_induction}, we identified cardiac spikes in the HR model. In another study, in preparation, we show that the phase diagram of the HR model closely resembles the KTz phase diagram (in this case, with $f(u)=tanh(u)$). Similarly to the KTz, in the HR model, BS evolves from CS when increasing the x-axis parameter of the phase diagram ($T$ in the KTz and $b$ in the HR). But in the HR diagram, as the parameter $b$ is increased, the plateau starts to oscillate regularly. These oscillations increase in amplitude, eventually replacing the plateau and generating the BS phase. Jiang and Hou~\cite{Jiang2015} used this "oscillating plateau" to create the spiral arms with the plateau and the substructures with the oscillations. Other papers with lattices of bursting HR cells presented similar substructures~\cite{Mainieri2005,Qin2014}. This is similar, albeit different from our results. The reason is that, in the KTz, CS and BS behaviors are mixed differently. Bursting (high amplitude oscillations) occurs immediately before or after the plateau or completely replaces it in a chaotic time series (as in Fig.\ref{fig:rede_x_I_card}-(b)). In the HR, bursting grows from regular oscillations in the plateau.

Despite its simplicity, the logistic KTz model was able to generalize and expand the results obtained previously with detailed conductance-based models, but without the computational burden of simulating dozens of variables and the adjustment of its several parameters. However, this comes with a price, reducing the biophysical plausibility and making it less clear how to translate the results into the experimental setting. The logistic KTz is a relatively new model, made with the general purpose of simulating neurons and cardiac cells, and was used in its current state. Future works could improve the biophysical plausibility of simulations by characterizing specific cardiac properties in the model, as the restitution curves~\cite{Clayton2010,Alonso2016}. The parameters could be tuned, likely venturing outside the present phase diagram, to make the cardiac plateaus longer and more robust to the high amplitude oscillations in the current, allowing the use of higher couplings. At the network level, different topologies and configuration that more closely resemble the anisotropy of the cardiac tissue could be adopted to verify its effect in the propagation properties. The \textit{torsades de pointes} observed in the activity is a significant result and requires further investigation, confirming its presence with a pseudoelectrocardiogram~\cite{Gima2002} (as performed by Sato et al.~\cite{Sato2009}) and understanding how it can manifest in a completely different sampling of the network. A pseudoelectrocardiogram would also allow investigation of other features in the simulations, as the long QT syndrome.

Polymorphic ventricular tachycardias like \textit{torsades de pointes} are life threatening arrhythmias. Studying these arrhythmias with quantitative details in simple descriptions, but retaining cellular characteristics like EAD, it is a promising topic for investigations made possible by a map-based approach. This is one of the very few studies to date of cardiac dynamics with map-based models and we have provided a throughout description on how to apply these equations to simulate arrhythmias. Currently, we are working into adapting the procedures described here for the neural setting and the study of the spiral waves found in the brain. 

\section*{Acknowledgments}
R.V.S. thanks the partial financial support from FAPESC and CAPES. This study was financed in part by the Coordenação de Aperfeiçoamento de Pessoal de Nível Superior - Brasil (CAPES) - Finance Code 001.

\section*{Supplementary Information}
Codes to reproduce the results and videos of the simulations are available at \url{https://github.com/rafaelvste/Reentry_CML_KTz}.


\end{document}